\documentclass[twocolumn]{aa}
\usepackage{graphicx}
\usepackage{txfonts}
\begin{document}
   \title{Upper limits on CO 4.7 $\mu$m emission from disks around five Herbig Ae/Be stars
   \thanks{Based on observations collected at the European Southern
Observatory, Chile (program ID 072.C-0535)}}
\titlerunning{CO 4.7 $\mu$m emission from disks around Herbig Ae/Be stars}

%   \subtitle{I. Overviewing the $\kappa$-mechanism}

   \author{A. Carmona
          \inst{1,3}
          \and
          M.E van den Ancker \inst{1}
	  \and
	  W.-F. Thi \inst{2}
          \and
          M. Goto \inst{3}
          \and
          Th. Henning \inst{3}
          }

   \offprints{A. Carmona\\
              \email{acarmona@eso.org}}
             
   \institute{European Southern Observatory, 
              Karl Schwarzschild Strasse 2 , 75748 Garching bei M\"unchen, Germany
         \and 
              Astronomical Institute, University of Amsterdam, Kruislaan 403, 
              1098 SJ  Amsterdam, The Netherlands 
            % \email{c.ptolemy@hipparch.uheaven.space}
            % \thanks{The university of heaven temporarily does not
            %         accept e-mails}
         \and
	     Max Planck Institute for Astronomy, K\"onigstuhl 17, 69117 Heidelberg, Germany
             }

   \date{March 11, 2005}

   \abstract{
     We present the results of medium-resolution spectroscopy
     of five nearby Herbig Ae/Be stars at 4.7 $\mu$m: 
     UX Ori, HD 34282, HD 50138, V380 Ori, HK Ori. 
     The goal was to search for CO fundamental ro-vibrational emission. 
     None of the targets show CO features, either in absorption 
     nor in emission. 
     We derive a $5\sigma$ upper limit of $< 10^{-12}$ cm$^{-2}$ to the column 
     density of hot CO (T $\approx$ 1500 K) in the sources. 
     These upper limits are considerably lower than 
     the values of Herbig Ae/Be stars for which warm and hot CO emission has been reported.
     The non-detection of CO $\nu$=1\--0 emission in these five targets 
     suggest that Herbig Ae/Be stars are not a homogeneous group with respect 
     to the structure of the gaseous disk and/or the amount of 
     CO in the inner 50 AU of their disks.

   \keywords{circumstellar matter -- ISM:molecules --
                planetary systems:protoplanetary disks                
               }
   }

   \maketitle
%
%************* INTRODUCTION ************

\section{Introduction} 
%**********TABLE of identifed lines *****************
\begin{table*}
\begin{minipage}[t]{\textwidth}
\caption{H\,{\sc i} lines detected in the 4.6--4.8 $\mu$m spectra of our targets}             % title of Table
\renewcommand{\footnoterule}{}  % to avoid a line before footnotes
\label{table:1}      % is used to refer this table in the text
\centering                          % used for centering table
\begin{tabular}{l l c c c c c}        % centered columns (4 columns)
\hline\hline                 % inserts double horizontal lines
                      &                & UX Ori     & HD 34282\footnote{No H\,{\sc i} lines were detected in the ISAAC spectra of HD 34282} & HD 50138   & V380 Ori   & HK Ori\\
\hline
Line                  &                & Pf $\beta$ & --        & Pf $\beta$ & Pf $\beta$ & Pf $\beta$ \\
Central Wavelength 
($\lambda$)           & [$\mu$m]       & 4.655      & --        & 4.656 & 4.655 & 4.655 \\
FWHM                  & [km s$^{-1}$]  & 255        & --        & 156   & 176   & 233 \\
Equivalent 
Width (EW)            & [$\AA$]        & $-$7.4       & --        & $-$11.7 & $-$12.7 & $-$5.1 \\
Line Flux 
(F$_{\lambda}$)       & [W m$^{-2}$]   & 2.3 $\times 10^{-16}$ & -- & 5.1 $\times 10^{-15}$ & 1.5 $\times 10^{-15}$ & 1.1 $\times 10^{-16}$\\
\\
Line                  &                & Hu $\varepsilon$         & -- & Hu $\varepsilon$         & Hu $\varepsilon$         & -- \\
Central Wavelength 
($\lambda$)           & [$\mu$m]       & 4.674                 & -- & 4.675                 & 4.673                 & -- \\
FWHM                  & [km s$^{-1}$]  & 60                    & -- & 142   & 116   & -- \\
Equivalent Width   
(EW)                  & [$\AA$]        & $-$3.2                  & -- & $-$6.7                  & $-$4.1                  & --\\
Line Flux 
(F$_{\lambda}$)       & [W m$^{-2}$]   & 1.0 $\times 10^{-16}$ & -- & 3.0 $\times 10^{-15}$ & 4.7 $\times 10^{-16}$ & --\\
\\
\hline                                   %inserts single line
\end{tabular}
\end{minipage}
\end{table*}
%*************************************************************
%\input{table1.tex}
%\input{introduction.tex}
%******************* INTRODUCTION ************************
Circumstellar disks around pre-main sequence stars are 
the likely place for planet formation.
Present theoretical models predict that planets 
are formed in the inner 50 AU of such disks. 
Terrestrial planets are thought to be made from hierarchical grouth of planetesimals 
(e.g. Ida \& Lin 2004).
Two major classes of models exist for the formation of gas giant planets:
a) the gravitational instabilities model (Boss 2004; Mayer et al. 2004),   
b) the core accretion model (Pollack 1996; Alibert et al. 2004; Kornet et al. 2002).           
The knowledge of fundamental disk properties
such as temperature and density profile,
geometry and dissipation time scales,
gas and dust composition, is  paramount to constrain
planet formation models.

Giant planets in our Solar System and giant extrasolar planets 
(Vidal-Madjar et al. 2004) are composed mainly of gas.
Current models predict that they are formed in the inner
50 AU of the protoplanetary disk (Ida \& Lin 2004; Alibert et al. 2004; Boss 2004).
However, the observational study of the gas in the planet forming region of the disks  
only started recently with the advent of space observatories 
(e.g. Bergin et al. 2004; Thi et al. 2001; Lecavellier des Etangs et al. 2003)
and high-resolution 
infrared spectrometers mounted on large aperture telescopes 
(e.g. Richter et al. 2002; Bary et al. 2002, 2003; Brittain \& Rettig 2002; 
Brittain et al. 2003; 
Najita et al. 2000, 2003; Blake \& Boogert 2004). 

Intermediate-mass Herbig Ae/Be stars (HAEBES) and
low-mass classical T Tauri stars are
pre-main sequence stars characterized by strong optical emission
lines (e.g. H$\alpha$) and infrared excess.
These properties are consistent with the idea that they are young stars
surrounded by a circumstellar disk.
The strong emission lines are interpreted as the signature
of gas accretion onto the central star, and the infrared excess
as the emission of the warm dust in the disk
(see reviews by Waters \& Waelkens 1998 and Bertout 1989).
These disks have been imaged from the NIR to the (sub-)millimetre
(e.g Grady et al. 2004; Mannings \& Sargent 1997, 2000; Henning et al. 1998). 
Nearby HAEBES and T Tauri stars
are natural laboratories for studying the process of planet formation.

Pure rotational CO emission from circumstellar disks is commonly seen 
at millimeter and submillimeter wavelengths 
(e.g. Qi et al. 2004; Ceccarelli et al. 2002; Mannings \& Sargent 2000; Thi et al. 2004).
However, these wavelengths are only sensitive to the cold gas (T $< 50 K$) 
in outer regions of the disk (R $> 50$ AU).

Theoretical models and recent observational evidence suggest that
in the inner 50 AU, temperatures can be relatively high 
($ T \geq 150 K$)
(Willacy et al. 1998; Dullemond et al. 2001; Markwick et al. 2002; Kamp \& Dullemond 2004).
Under these conditions, molecules such as water and CO are in the gas phase,
and a rich emission spectrum from their rotational and vibrational 
transitions is expected.
However, the relatively large column densities in the inner 50 AU 
(1500 g cm$^{-2}$ at 1 AU for the minimum-mass solar nebula) 
could preclude the observability of such transitions, 
because the disk is optically thick in the continuum (Najita et al. 2003) . 

Nevertheless, in some circumstances, parts of the inner circumstellar disk
could be optically thin, and warm CO emission becomes detectable.
Carr et al. (2001), Brittain \& Rettig (2002) and  Rettig et al. (2004) 
detected $\nu = 1\--0$ emission of warm CO and suggested   
that it originates in a disk gap or in a inner low-density region;
Najita et al. (2000, 2003) and Blake \& Boogert (2004) reported the detection 
of CO fundamental ro-vibrational band, and postulated that warm CO emission is produced 
in a disk's atmosphere when the stellar radiation of the 
host star induces a temperature inversion 
(disk atmosphere hotter than the disk mid-plane);
Brittain et al. (2003) found $\nu = 1\--0$ emission of hot and warm CO in AB Aur and 
suggested that it is produced by IR pumping (resonant scattering). 
The same authors reported the detection of $\nu = 2\--1$~and~$3\--2$ emission of CO in HD 141569, 
and pointed out that it originates by UV pumping in the inner rim of the disk.   

This paper describes our search for CO fundamental ro-vibrational emission 
from a selected number of Herbig Ae stars known to be surrounded by a disk.
Meeus et al. (2001) devised a classification scheme for the spectral energy distributions
(SED) of Herbig Ae/Be stars. 
In their classification scheme, group I sources 
have a SED that is strongly double-peaked in the near to mid-infrared, whereas
the SED of group II  sources
can be described by a power-law at those wavelengths.
Dullemond (2002) has identified this empirical classification with two 
different disk geometries: flared (group I) and self-shadowed (group II).
We observed objects belonging to both groups.
UX Ori, V380 Ori, HK Ori (group I) are likely to have a flared disk, 
and HD 50138 and HD 34282 (group II) a self-shadowed disk. 
We used ISAAC at the VLT at R$\sim$10,000 to search for the CO emission
band at 4.7~$\mu$m. We report here our non detection in all the objects observed in 
April 2004. We discuss how the stringent upper limits derived from our deep search set 
constrains the physical properties of the observed disks.

The paper is organized as follows.
In \S2 the observational set-up and data reduction procedure are described,
in \S3 the spectra obtained and the estimation of the upper limits 
to the hot (T $\approx$ 1500 K) CO column density are presented, 
finally in \S4 the meaning of our results in the context of the structure of the
disk is discussed. 

%************** OBSERVATIONS ***********

\section{Observations} 
%\input{table2.tex}
%***************TABLE of STATISTICS **************************
\begin{table*}
\begin{minipage}[t]{\textwidth}
\caption{Summary of CO $\nu$ = 1--0 emission charateristics of our sample stars.}             % title of Table
%\centering
\renewcommand{\footnoterule}{}  % to avoid a line before footnotes
\label{table:2}      % is used to refer this table in the text
\centering                          % used for centering table
\begin{tabular}{l l r r r r r r r}        % centered columns (4 columns)
\hline\hline                 % inserts double horizontal lines
 & & UX Ori   &  HD 34282 & HD 50138 & V380 Ori & HK Ori & AB Aur
\footnote{Values for AB Aur are listed for reference (Brittain et al. 2003)} \\
\hline 
Distance \footnote{For HD 34282, HD 50138 and AB Aur the distance is estimated using Hipparcos data.  
UX Ori, V380 Ori and HK Ori have been assumed to be associated
with the Orion OB1a (UX Ori) and OB1c (V380 Ori, HK Ori) star-forming regions. }
                  & pc & 400 & 164 & 290 & 500 & 500 & 144\\
Average Intensity & [$10^{-13}$ W m$^{-2}$ $\mu$m$^{-1}$] & 3.1 & 1.8 & 44.7   & 10.0    & 2.2     & 14.7 \\
5$\sigma$         & [$10^{-13}$ W m$^{-2}$ $\mu$m$^{-1}$] & 0.6 & 1.3 & 9.0    & 1.1     & 0.5     & -- \\
S/N               &                                       & 25  & 7   & 25     & 45      & 23      & -- \\  
Instrument FWHM   & [$10^{- 4}$ $\mu$m]                   & 4.7 & 4.7 & 4.7    & 4.7     & 4.7     & -- \\
Line Flux upper-limit  & [$10^{-17}$ W m$^{-2}$]  & $<$ 2.9 & $<$ 5.9 & $<$ 42 & $<$ 5.2 & $<$ 2.2 & 8.6 \\
scaled $g_{1 \-- 0}$   & pho mol$^{-1}$ s$^{-1}$  & 1.8     & 0.2     & 13.6   & 9.0     & 2.0     & 1.1 \\ 
Column density of 
hot CO (1500K)   & [$10^{12}$ cm$^{-2}$]          & $<$ 4.1 & $<$ 89 & $<$ 8.0 & $<$ 1.5 & $<$ 2.9 & 20  \\
\hline                                                        
\end{tabular}
\end{minipage}
\end{table*}
%*************************************************************************************

%\input{observations.tex}
Medium-resolution spectra in the 4.6 -- 4.8~$\mu$m range 
were obtained for five 
Herbig Ae/Be stars (UX Ori, HD 34282, HD 50138, V380 Ori and HK Ori) 
between the 2$^{nd}$ and 20$^{th}$ of April 2004 
using ISAAC  
at the First Unit Telescope ANTU of the ESO-VLT 
at Cerro Paranal, Chile. 
These targets were selected to maximize the shift in velocity 
between the expected position of the CO emission in the targets
and the telluric CO absorption at the time of observation.
A medium-resolution grating and the narrowest 
available slit (0.3") were used to provide spectra
of resolution 10,000.
The slit has been oriented in the North-South direction.
Sky background was subtracted by chopping each exposure 
by 15 '' in the direction of the slit.
Asymmetrical thermal background of the telescope was subtracted  
by nodding the telescope by 15''.
%The sequence of raw frames has been obtained employing the standard A B B A technique,
%were A and B represent the two nodding positions.
In each nodding position the telescope has been randomly jittered in order
to record the raw spectra in different regions of the detector,  
minimizing the influence of bad pixels.
In order to correct telluric absorption and obtain absolute flux calibration,
spectroscopic standard stars were observed the same night at airmasses 
close to that of the science targets. 
Dome flat fields were obtained at the beginning and at the end of each observing night.

\subsection{Data Reduction} 
Chopped raw frames of each nodding position were first corrected for the detector 
non-linearity. 
Since half-cycle frames were not recorded, 
the median value of each frame was subtracted to obtain an approximation 
to the half-cycle frame intensity subject to non-linearity.
The residual raw frame was corrected for non-linearity using the expression
\begin{center}
$F_{c} =  f_{rr} + 1.05 \times 10^{-6} \times f_{rr}^2 + 0.85 \times 10^{-10} \times f_{rr}^3$
\end{center}  
in which $F_{c}$ refers at the corrected frame, and $f_{rr}$ is the residual raw frame
\footnote{Values taken from the ISAAC data reduction manual, see 
www.eso.org/instruments/isaac/\#Documentation}. 
Once this correction for detector non-linearity is made, the median previously 
subtracted is added back to the frame. 

The second step in the data reduction procedure is to correct all the frames 
for tilt in the spatial direction using Ar Xe arc 
lamp frames taken the same day of the observations. A second order polynomial 
deduced from the arc file is used for this correction. 
Individual exposures are corrected for differences in the detector sensitivity 
by dividing them by normalized dome flat fields. 

Flat-fielded frames were combined using the Eclipse\footnote{www.eso.org/eclipse} 
jittering procedure. 
The procedure consists in classifying the set of data in groups having the
nodding sequence form ABBA. 
In each group, the frames in the A position and B position are averaged,
the averaged frames are subtracted from each other, divided by 2, and a combined frame is generated 
($F_{combined}= (\bar{A} - \bar{B})/2$). 
The ensemble of combined frames, one for each ABBA group, are stacked in one single final
2D frame. During this process, combined frames are shifted in such a way that 
the addition of all the spectra is located in the center of the the final frame.
This shifting procedure optimizes the subtraction of 
sky and telescope background emission. 

One-dimensional spectra are extracted from the 2D frame, averaging the pixel counts
in the PSF direction. These spectra are then divided by the exposure time defined
by
\begin{center}$E_{time}[s]=\rm{DIT}\times \rm{NDIT}\times 2\times \rm{Chopcycles}$\end{center}
Here DIT is the Detector Integration Time (1.84 s in our case),
NDIT is the number of DIT and Chopcycles is the number of chopping cycles (8) per DIT.  

The one-dimensional spectrum of the standard star is corrected for the 
differences in airmass and air pressure with the science target using,
\begin{center}
%$I_{STD_{corrected}} = I_{0}\; \exp{\left(-\tau \; \frac{Target_{average\_airmass}}{STD_{average\_airmass}} \frac{Target_{average\_airpress}}{STD_{average\_airpress}}\right)}
$I_{STD_{corrected}} = I_{0}\; \exp{ \left(
-\tau \;
\frac{\displaystyle \bar{X}_{TARGET}}{\displaystyle \bar{X}_{STD}} \; 
\frac{\displaystyle \bar{P}_{TARGET}}{\displaystyle \bar{P}_{STD}} \right) }
$
\end{center}
Here $\bar{X}$ is the average of the airmass and 
$\bar{P}$ is the average of the air pressure.
The continuum $I_{0}$ has been defined as uniform with a value equal to the mean plus 
three standard deviations
($I_{0}= \bar{I}_{\rm obs} + 3\sigma_{I_{\rm obs}}$). 
The optical depth $\tau$ has been calculated from
the measured data using $\tau = \ln{\left(I_{obs}/{I_{0}}\right)}$.
 
Employing a synthetic model of the earth atmosphere,
the extracted one-dimensional spectra of the target and the standard star
were subject to a precise wavelength calibration involving 
a correction for differences in the spectral resolution of both stars (Goto et al. 2003).

Finally, the science target spectrum is flux calibrated and corrected for atmospheric 
absorption by dividing it by the corrected spectrum of the standard star.
Absolute flux calibration is obtained by multiplying the telluric corrected spectra by a 
theoretical SED model (Kurucz 1991) of the standard star. 

We estimate the uncertainty in the flux calibration of our spectra to be 
between 10\% and 20\%. 
The principal source of noise in the spectra is due to imperfections in 
the correction for telluric absorption.
The 4.7 $\mu$m band is particularly challenging in this aspect given the 
presence of strong CO absorption in the earth's atmosphere. 
Although the standard stars were observed with airmasses close to those of 
the science targets,
and the extracted spectra were corrected for differences in airmass, air pressure and spectral
resolution, in the regions of poor atmospheric transition the correction 
for the telluric absorption is not perfect.
The noise present in the spectra is mostly due to systematic errors 
in this correction. 

%************** RESULTS *****************

\section{Results}

%*** FIGURE SPECTRA ***
\begin{figure*}
\centering
\includegraphics[angle=0,width=0.85\textwidth]{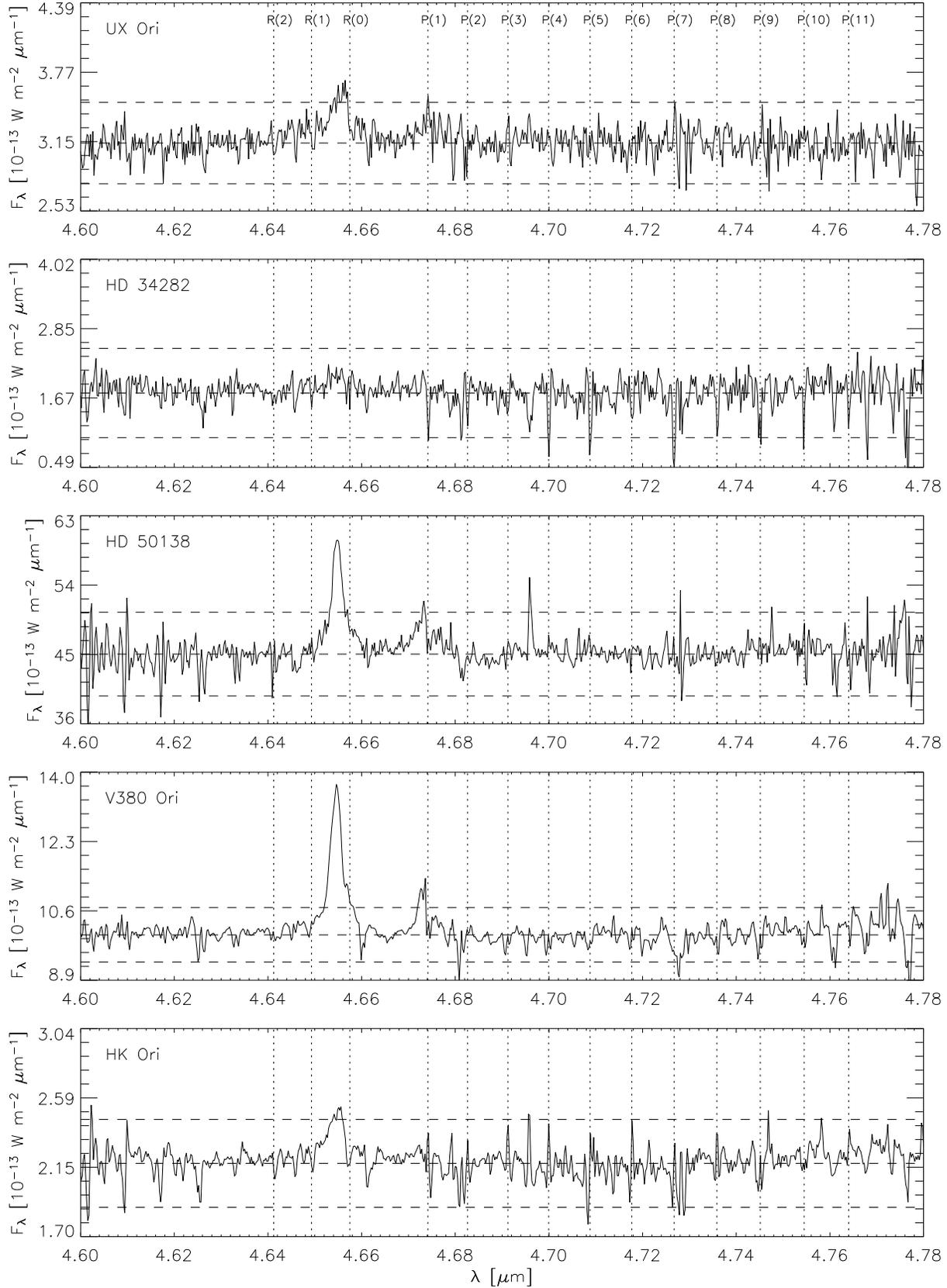}
      \caption{
       ISAAC spectra of our five target stars. 
       The strong emission seen at 4.65 $\mu$m and 4.67 $\mu$m are
       the Pf $\beta$ and Hu $\varepsilon$ recombination lines of H\,{\sc i}. 
       The noise in the spectra is mostly due to systematic 
       errors in the telluric correction in this region of poor atmospheric
       transmission. Vertical dotted lines show the CO $\nu = 1\--0$ transitions. 
       Horizontal dashed lines show the 3$\sigma$ limits of the intensity.  
       }
         \label{ISAACspectra}
\end{figure*}
%**********************

%\input{results.tex}
Flux calibrated spectra of the five Herbig Ae/Be stars are presented in Figure 1.
The $\nu$ = 1--0 ro-vibrational band of CO was not detected in any 
of our targets, neither in absorption nor in emission.
The only notable feature in all the spectra (except HD 34282)
are recombination lines of H {\small I}:
Pf${\;\beta}$, Hu${\;\varepsilon}$. 
A summary of the H\,{\sc i} observations is presented in Table 1. 
In the case of HD 293782 and HK Ori the Pf$\,{\beta}$ line shows 
a strong red/blue asymmetry.
This phenomenon is likely associated with infalling gas (Najita et al. 1996).  
HD 50138 and V380 Ori show symmetric Pf$\;{\beta}$ profiles. 
In these two systems the Hu${\;\varepsilon}$ emission line has a 
strong blue/red asymmetry and extended blue wings.
This line most likely originates in a similar region as the balmer lines
of H\,{\sc i}, which are believed to come from a wind either
from the disk or the central star (see Bouret \& Catala 1998, 2000).
 
We derived upper limits to the flux in the CO emission line 
taking a conservative 5$\sigma$ (standard deviation)
of the intensity in the continuum ($\sigma_{I{_{obs}}}$).   
Given that the minimum FWHM of the instrument is 
$4.7 \times 10^{-4}$ $\mu$m, we obtained upper limits 
ranging from $2.2 \times 10^{-17} \;\rm{to}\; 5.9 \times 10^{-17}$ W~m$^{-2}$ for 
UX Ori, HD 34282, HK Ori, V380 Ori,
and $4.2 \times 10^{-16}$ W~m$^{-2}$ for HD 50138 (see Table 2).        

From the upper limit to the CO flux, and supposing that CO 
is excited by infrared fluorescence (resonant scattering), 
we estimated the upper limit to the column density of hot CO using 
Brittain et al. (2003)\footnote{originally in DiSanti et al. (2001)}:
\begin{center}
$N= \frac{\displaystyle 4 \pi F_{line}}{\displaystyle \Omega\;g_{line}\;h\,c\,\tilde{\nu}}$
\end{center}       
Here $F_{line}$ is the $\nu = 1\--0$ line flux,
$\Omega$ is the solid angle,$h\,c\,\tilde{\nu}$ is the energy per photon, 
and $g_{line}=P_{J}\,g_{1-0}$ is the fluorescence efficiency for the $1\--0$
ro-vibrational transition by IR pumping.

Brittain et al. (2003) estimated a  
$g_{1-0}$ at 0.6 AU of 1.1 photons molecule$^{-1}$s$^{-1}$ for AB Aur, 
and calculated a column density of observed hot CO of $2 \times 10^{13}$ cm$^{-2}$.
Given that UV pumped ($\nu = 2\--1$) transitions were not detected 
in the wavelength range observed (4.6 \-- 4.8~$\mu$m), 
and assuming that our observed HAEBES have an inner disk
configuration similar to AB Aur, we scaled the $g_{1-0}$ coefficient
for each star based on the IR luminosity ($L_{\,\rm{IR}\,_{\star}}$) 
at 4.7$\mu$m and the distance ($d_{\star}$) employing
\footnote{AB Aur is a 144 pc and has a 4.7 $\mu$m intensity 
of 1.47 $\times$ $10^{-12}$ W m$^{-2} \mu$m$^{-1}$.}
\begin{center}
$g_{1-0\;\star}= g_{1-0_{\rm{AB\,Aur}}} \; 
\frac{\displaystyle d^{2}_{\star} \times L_{\,\rm{IR}\,_{\star}}}
{\displaystyle d^{2}_{\rm{AB\,Aur}} \times L_{\,\rm{IR}_{\rm{AB\,Aur}}}} 
$ 
\end{center}       
With this estimate of $g_{1-0}$ for each of our targets 
and the upper limit for the line flux we calculated the hot CO column density.  
Our results are summarized in Table 2. 
We find upper limits to the column density of hot CO ranging from 
$1.5 \times 10^{12} \;\rm{to}\; 8.0\times 10^{12}$ cm$^{-2}$. 
In the case of HD 34282 the upper limit calculated is $8.9 \times 10^{13}$ cm$^{-2}$.
However, this is not a representative limit since the signal to noise of the
spectra is low (S/N=7). 
 
%****************** DISCUSSION ***********************

\section{Discussion} 
Najita et al. (2003), Brittain et al. (2003), 
Blake \& Boogert et al. (2004), and Rettig et al. (2004)
showed how  the study of warm and hot  CO emission 
could be used as a powerful tool for probing disk surfaces and for assessing  
the density and temperature profiles
of the inner 50 AU of the circumstellar disk independent of SED models.
These authors reported the detection of the $\nu = 1 \-- 0$ transition of CO 
in several Herbig Ae/Be stars. 

Our estimates of the upper limit to the CO line fluxes are within the 
range of the detections of CO $\nu = 1\--0$ emission fluxes
measured.
% by these authors.
For example in AB Aur Brittain et al. (2003) observed a 
CO $R(1)$ line flux of  $8.6 \times 10^{-17}$ W~m$^{-2}$ in AB Aur
and Blake \& Boogert (2004) detected a CO $P(2)$ line flux of 
$10.4 \times 10^{-17}$ W~m$^{-2}$ in the same star.
The upper limits to the hot CO column density calculated 
(except for HD 34282 that presented a low S/N spectrum)
are well below the $2 \times 10^{13}$ cm$^{-2}$ 
value reported for AB Aur by Brittain et al. (2003). 
From the upper limits deduced from our observations,
we conclude that if our targets would had equal column densities of hot CO, 
or had presented CO $\nu =1\--0$ 
emission fluxes of the magnitude reported in the literature, 
we would have detected this emission within our ISAAC data. 

Previous studies suggested that CO emission is 
produced by UV pumping or by IR fluorescence (resonant scattering).
The nature of the excitation mechanism is revealed from the CO transitions
present in the spectra. 
The $\nu = 2\--1\;$and$\;3\--2$ emission bands indicate that 
the principal source of excitation is UV pumping.
On the other hand, the $\nu = 0\--1\;$ band indicates that the excitation
mechanism is thermal (i.e. IR resonant scattering). 
Our upper limits are small enough 
that they are in the detection range of the
$\nu = 2\--1\;$and$\;3\--2$~ emission band 
previously reported in the Herbig star HD 141569 
by Brittain et al. (2003), $F_{3\--2}=$1.1$\times 10^{-17}$ W~m$^{-2}$). 
The non-detection of UV excited CO transitions within our data
suggests that UV radiation is scattered by the dust in the disk's inner rim
and that the CO is probably shielded.
However, 
the inner part of the disk (R $< 50$ AU) could also be optically thick in the continuum,
precluding the observability of such transitions.

Meeus et al. (2001) in their study of the spectral energy 
distribution (SED) of a large sample of Herbig Ae/Be stars 
suggested that they could be classified in two groups according to 
the SED shape. 
Detailed modeling of the dust emission 
by Dullemond (2002) found that
the 3 $\mu$m bump observed in many SEDs could be interpreted as evidence for
the existence of a dusty ''puffed-up'' inner disk rim, 
and  that the Meeus et al.  classification could be understood as 
corresponding to two inner disk geometries: flared disk, and self-shadowed disk.

Using this geometrical classification, Brittain et al. (2003)
interpreted the two temperature CO $\nu = 1\--0$ profiles 
deduced from the rotational diagrams of AB Aur in the following way. 
They suggested that the hot (1500 K) CO originates just beyond the
``puffed-up'' inner rim by IR resonant scattering at 0.6 AU, 
and that the cold CO (70 K) is emitted by 
IR fluorescence in the surface of the flared disk at 8 AU.
 
Following this classification method, 
our targets fall in both groups of disks:
UX Ori, V380 Ori, HK Ori are likely to have a flared disk, 
and HD 50138 and HD 34282 a self-shadowed disk. 
Their SEDs indicate that all of
them should have a ``puffed-up''~inner rim. 
The non-detection of $\nu = 1\--0$ emission in our targets suggests that 
there is no strong correlation between the disk geometry and CO emission properties.

Recent work by Rettig et al. (2004), shows the detection of 
optically thin warm (T $\approx$ 140 K) CO emission in the classical T Tauri 
star TW Hydrae. 
These authors ascribed the non detection of hot CO (T $>$ 1000 K) as evidence 
for a cleared inner disk region out to a radial distance of $\approx$0.5 AU, 
and proposed that the warm CO emission is produced in a dissipating gaseous disk.
In view of this interpretation, 
we could conclude that the non-detection of hot CO in our targets is consistent 
with the idea that our stars cleared of gas in their inner gaseous disks. 
However, the presence of strong emission lines in the optical spectra of HAEBES
testifies to the occurrence of accretion. 
Both results could be compatible under a scenario where accretion and 
clearing are episodic and that a replenishment mechanism is active.
%Nevertheless, TW Hya is relatively cool star and it is possible 
%that the flux from the star is not sufficient to exict enough gas
%at T$>$1000K to be detectable. 

In summary, 
the non-detection of CO $\nu = 1\--0$ emission in these five targets suggests that, 
{\it despite the relative similarity of the dust emission in Herbig Ae/Be stars,
they are not a homogeneous group with respect to gas emission, and in 
particular to hot and warm CO emission}.
This heterogeneity could be interpreted as the result of
(1) differences in the structure of their inner gaseous disk
leading to differences in the inner disk CO temperature, or
(2) true differences in the amount of warm CO, either due to chemical 
effects (selective CO depletion implying a variable CO/H$_{2}$ ratio in the disk) 
or due to variations in the amount of bulk gas in the inner 50 AU (assuming a constant CO/H$_{2}$ ratio).
The last scenario is particularly intriguing.
True variations in the the bulk of H$_{2}$ in the inner disk 
are consistent with the idea that in these stars giant planets have 
already formed.
Unfortunately, the present data do not allow us to confirm such
a tantalizing hypothesis. 
Future instruments with improved spectral resolution like CRIRES at the VLT
will help us to address such ideas quantitatively.

%______________________________________________________________

\begin{acknowledgements}
We would like to thank C.P. Dullemond and A. Bik for several fruitful discussions.
The authors wish to thank the ESO-VLT staff at Paranal observatory 
and in Garching that performed the service-mode observations presented in this paper.
This research has made use of the SIMBAD database,
operated at CDS, Strasbourg, France.      
\end{acknowledgements}

%______________________________________________________________

%______________________________________________________________

\end{document}